\documentstyle[12pt]{article}
\newcommand{\beq}{\begin{equation}}
\newcommand{\eeq}{\end{equation}}

\newcommand{\dd}{D\hspace{-.65em}/}
\newcommand{\nn}{\partial\hspace{-.55em}/}

\newcommand{\r}{\varphi}

\newcommand{\e}{\bar{\eta}}
\newcommand{\p}{\bar{\psi}}
\newcommand{\s}{\varepsilon}

\def\op{operator}
\def\dop{Dirac operator}
\def\tly{topologically}
\def\tl{topological}

\def\inst{instanton}

\def\gl{global}
\def\an{anomaly}
\def\f{field}

\def\nl{normalizable}
\def\nty{normalizability}
\def\zm{zero mode}
\def\lc{level crossing}
\def\mm{mass matrix}
\def\igty{integrability}
\def\cn{condition}
\def\con{configuration}
\begin{document}
\begin{titlepage}
\begin{flushright}
NBI-HE-93-51 \\
September 1993\\
\end{flushright}
\vspace{0.5cm}
\begin{center}
{\large {\bf The $SU(2)$ Global Anomaly Through Level Circling}}\\
\vspace{1.5cm}
{\bf Minos Axenides}
\footnote{e-mail:axenides@nbivax.nbi.dk}\\
\vspace{0.4cm}
{\em The Niels Bohr Institute\\
University of Copenhagen, 17 Blegdamsvej, 2100 Copenhagen, Denmark}\\
\vspace{0.4cm}
{\bf Andrei Johansen}
\footnote{e-mail:johansen@lnpi.spb.su}\\
\vspace{0.4cm}
{\em The St.Petersburg Nuclear Physics Institute\\
 Gatchina, St.Petersburg District, 188350 Russia}\\
\vspace{0.4cm}
{\bf Holger Bech Nielsen}
\footnote{e-mail:hbech@nbivax.nbi.dk}\\
\vspace{0.4cm}
{\em The Niels Bohr Institute\\
University of Copenhagen, 17 Blegdamsvej, 2100 Copenhagen, Denmark}\\

\end{center}
\begin{abstract}
We discuss a novel manifestation of the $SU(2)$ global anomaly in an
$SU(2)$ gauge theory with an odd number of chiral quark doublets
and arbitrary Yukawa couplings. We
argue that the massive 4-dim.($D=4$) Euclidean Dirac operator is nonhermitean
with its spectrum of eigenvalues $(\lambda,-\lambda)$ lying in pairs in
the complex plane. Consequently the existence of an odd number of
normalizable zero modes of the 5-dim.($D=5$) massive Dirac operator is
equivalent to a fermionic level exchange phenomenon, level ``circling'',
under continuous topologically nontrivial deformations of the external
gauge field. More generally global anomalies are a manifestation of
fermionic level ``circling'' in any $SP(2n)$
gauge theory with an odd number of massive fermions in the spinor
representation
and arbitrary Yukawa couplings.

\end{abstract}
\end{titlepage}
\newpage

\section{Introduction}
\setcounter{equation}{0}

The phenomenon of fermion level crossing and its relation to
the existence of normalizable fermionic zero modes of the \dop \ in
the presence of \tly \ non-trivial gauge fields is well known \cite{sc}.
It has been extensively studied and firmly established for the case of
massless fermions.

The $SU(2)$ global anomaly \cite{witten}
is also related to the phenomenon of
level crossing of the $D=4$ Dirac operator.
This eigenvalue flow
corresponds to the existence of a \zm \ for an appropriately defined
$D=5$ \dop \
in the presence of an external topologically
non-trivial gauge field. As a result the fermionic path integral
is not gauge invariant and the theory is not self-consistent.

In the realistic electro-weak theory fermions become massive
due to the Higgs mechanism and the presence of Yukawa interactions
with the Higgs field being a non-vanishing constant at spatial infinity.
At first appears surprising the occurence of \lc \ in the
spectrum of massive fermions.

We, however, offer an intuitive
argument for its existence in this case.
Let us first consider the $D=5$ \dop \
in an external \tly \ nontrivial gauge \f .
Such a configuration is \tly \ guaranteed by the fact that
$\pi^4(SU(2))=Z_2$ \cite{hu}.
In our case where Yukawa interactions are present it is tacitly taken that
Higgs \f s are suitably included as external \f s.
As a consequence of our construction we take at $\bf R^5$ infinity our gauge
\f s to be a pure gauge $A_{\mu}\propto U^{-1}\partial _{\mu}U$ with
$M\propto U$ being the fermion \mm .
Here $U$ is a unitary matrix, a noncontractible map from $S^4$ into
$SU(2)$ on a sphere at $\bf R^5$ infinity.
It means that one can continuously deform it to the identity everywhere
on the sphere except for the region near a single point where $U\neq 1$
(a singularity).
As the radius $R$ of the sphere takes value from infinity to zero we form
a line in the neighbourhood
of which the \tl \ non-triviality is concentrated
and the external \f \ changes very rapidly.
Therefore at $R=0$ we should have either zero or singularity.
By continuity of the \mm \ the latter is impossible, and hence we should
expect the existence of zero in $M$ with \lc \ to occur.

We have previously studied\cite{mah}
the effect of Yukawa interactions
for the level crossing phenomenon in the case of Witten's global
anomaly.
However our analysis was restricted to the symmetric case of equal
up and down quark masses in an $SU(2)$ doublet.
Herein we present an ananlysis for the more general and realistic
case of arbitrary Yukawa couplings and nondegenerate up and down
quark masses. The main difference with the symmetric case we find to be
in the nonexistence of an appropriate hermitean generalization for the
$D=4$ and $D=5$ Dirac operators. They in fact turn out to be
nonhermitean with their spectra of eigenvalues complex. In the present
work we formulate generalization of the zero mode index theorems and
demonstrate the $SU(2)$ global anomaly to be a manifestation of a
generalization of the fermionic level crossing phenomenon. Under
continuous topologically nontrivial deformations of the $SU(2)$ external
gauge field the fermionic eigenlevels of the $D=4$ Dirac operators lying
in the complex plane exchange their sign with one another
$(\lambda,-\lambda)$ pairwise by `circling' around the origin
$\lambda=0$. The level circling is shown to be a result of the existence
of an odd number of normalizable zero modes of the $D=5$ massive Dirac
operator. Moreover our arguments can be shown to trivially
generalize to the
case of $SP(2n)$ groups.

\section{Global Anomaly for Massive Fermions}
\setcounter{equation}{0}

It is well known that $SP(2n)$ gauge theories have no
local anomalies. Moreover when coupled with an odd
number of Weyl fermions they possess
a global anomaly and become self-inconsistent.
The simplest such example is an $SU(2)$ gauge
theory with one fermionic Weyl doublet.
This model was previously considered
in some detail for the case of massless fermions \cite{witten}.
In this section we extent Witten's arguments in order to take into
account the presence of Yukawa interactions.
We will show that the Atiyah-Singer index theorem
mod 2 is sufficient for the presence of the \an \ as a consequence
of the existence of a massive normalizable \zm \ of the $D=5$ \dop .

Let us first sketch Witten's arguments \cite{witten}.
The partition function of
the euclidean version of the model of a doublet of
massless fermions coupled to an
$SU(2)$ gauge field reads as follows
\beq
Z = \int D\psi_L D\psi^+_L \int DA_{\mu} exp(-\int d^4 x
[(1/2g^2)Tr F_{\mu\nu}^2 + \psi^+_L i\nabla \hspace{-.65em}/ \psi_L]).
\eeq
There  $A_{\mu}$ is an $SU(2)$
gauge field, $\psi_L$ is a left-handed Weyl fermion
doublet,
$g$ is the gauge coupling constant, $\nabla \hspace{-.65em}/
 = \nabla_{\mu}\gamma_{\mu}$
is a \dop \ restricted to act on a Weyl doublet.
The fermionic
part of the integral eq.(2.1) is ill defined.
However it can be formally integrated as the square root of
a functional integral over Dirac fermions.
As such it implies the doubling of the
fermionic degrees of freedom
from one to two Weyl left-handed doublets.
Because the $1/2$ representation of $SU(2)$ is pseudoreal a left-handed
doublet can be mapped to the right-handed one.
A theory with two left-handed Weyl doublets
is thus equivalent to a vector-like one with a single Dirac doublet.
The fermionic functional integral is given by
$det (i\nabla \hspace{-.65em}/ )$
and it is well defined.
Then we formally have that
\beq
\int D\psi_L D\psi^+_L exp \int \psi^+_L i\nabla \hspace{-.65em}/
\psi_L =
(\det i\nabla \hspace{-.65em}/ )^{1/2}.
\eeq
The sign of the
square root is ill defined. As a way out Witten defines
the root in eq.(2.2) as the product of all positive eigenvalues of
a \dop as a start and then continuous analytically.

If for a given $A_{\mu}$ the sign in eq.(4.2)
is arbitrarily fixed as Witten showed
there always exists
a configuration $A_{\mu}^U$ that
can be reached continuously
from $A_{\mu}$ for which the fermionic determinant has an opposite
sign, i.e.
\beq
(det i\nabla \hspace{-.65em}/ (A_{\mu}))^{1/2} =
- (det i\nabla \hspace{-.65em}/ (A_{\mu}^U))^{1/2}.
\eeq
Here $A_{\mu}$ is taken to be the gauge transformed
\con \ of $A_{\mu}$.
This means that the partition function
\beq
Z = \int DA_{\mu} (det i\nabla \hspace{-.65em}/ )^{1/2}
exp (-(1/2g^2)\int d^4 x Tr F^2_{\mu\nu})
\eeq
vanishes due to the contribution with an opposite sign from $A_{\mu}$ and
$A_{\mu}^U$.
This comes about because as we continuously vary
the external field value from $A_{\mu}$ to $A_{\mu}^U$ an odd number of these
eigenvalues flow through zero switching their sign.
Such a \lc \ effect is a reflection of the existence of an odd number of
\nl \ \zm s for a properly defined $D=5$ \dop \ in a topologically
nontrivial gauge \f .
It can happen because of the nontrivial fourth homotopy group of $SU(2)$
\beq
\pi^4(SU(2)) =Z_2.
\eeq
The D=5 topologically non-trivial gauge field must accordingly
belong to the
nontrivial homotopy class in $Z_2$
and it is the interpolating \con \ between $A_{\mu}$ and $A_{\mu}^U$.
An appropriate $D=5$ \dop \ is actually an extension of the $D=4$
one that includes evolution in the fifth coordinate $\; t$.
In what follows we generalize Witten's argument for the case of Yukawa
interactions and the presence of fermion mass through the Higgs
mechanism.
In this context we will prove Witten's conjecture of the validity
of his arguments for this case too.
It is worthwhile to also stress that the $SU(2)$ \gl \ \an \
can also be understood as a manifestation of the existence of a local
\an \ for an $SU(3)$ gauge theory \cite{klinkhamer}.
{}From this point of view the generalization to the case of massive
fermions is of course straightforward.
This is due to the fact that the local \an \ is independent of Yukawa
couplings which are $SU(3)$ invariant \cite{fuj}.
It seems
nevertheless interesting to try to understand the issue in terms of
a level crossing phenomenon
for the $D=4$ Dirac \op .
This is the aim of what is to follow.
We do it by proving a generalization of the index theorem
mod 2 for the massive $D=5$ \dop .

We first consider the index theorem
mod 2 for a massless $D=5$ Dirac \op .
Instead of making
a unitary transformation to a real representation
for the $D=5$ fermions which transform under the
$O(4) \times SU(2)$  group \cite{witten} we introduce
\beq
D\hspace{-.65em}/ = (\gamma_5 \nabla_t + \nabla \hspace{-.65em}/ ).
\eeq
Here $\nabla_{\mu}$ is the usual covariant derivative,
$\nabla \hspace{-.65em}/ = \nabla_{\mu} \gamma_{\mu}$,
whereas $\nabla_t$
is the covariant derivative for the
fifth coordinate.
Eigenvalues and eigenfunctions are defined by the following equation
\beq
D\hspace{-.65em}/  \psi = i \lambda \psi ,
\eeq
where $\lambda$'s are real since $\dd$ is antihermitean.
The \op \
$D\hspace{-.65em}/ $
is real and antisymmetric in the following sense
\beq
C\epsilon \gamma_5 \dd^* C\epsilon \gamma_5 = \dd  ,
\eeq
\beq
C\epsilon \gamma_5 \dd^T C\epsilon \gamma_5 = - \dd.
\eeq
Here $C=i\gamma_2\gamma_0$ and $C\gamma_5$
are the usual $D=4$ and $D=5$ charge conjugation matrices,
$\epsilon$ is a $2\times 2$ antisymmetric matrix that acts on $SU(2)$
indices.
The reality \cn \ implies the pairing up of all the non-zero
eigenvalues.
More precisely
if $\dd \psi = i \lambda \psi$ then
\beq
\dd (C\epsilon \gamma_5 \psi^*) = -i \lambda C\epsilon \gamma_5 \psi^.
\eeq
The zero mode wave
functions $\psi_0$ (if they exist) can be chosen real
satisfying a Majorana-like \cn :
\beq
C\epsilon \gamma_5 \psi_0^* = \psi_0.
\eeq
This reality condition means that the number of zero modes
is a topological invariant modulo 2 since non-zero modes can
cross the zero level only in pairs when external fields vary smoothly.
The crucial point is that such a zero mode does exist in the external
field due to the index theorem modulo 2 \cite{atiyah} if
the external \f \ is topologically nontrivial.
The reality of the Dirac operator
$\dd$ guarantees
a pairing of its eigenvalues.
They correspond to the existence of the nontrivial homotopy class that
arises in the nontrivial topology $\pi^4 (SU(2)) = Z_2$.

Let us emphasize here that the generalization of the above
considerations to the case of $SP(2n)$ gauge groups is straightforward.
This is because the spinor representation of $SP(2n)$ group is
pseudoreal while the corresponding representation of
$O(5) \times SP(2n)$ group is real.
Actually all the formulas in this paper are valid for these groups too
with a change of the isotopic antisymmetric tensor into
a matrix $s$ of charge conjugation of the $SP(2n)$ group
which is given by a $2n \times 2n$ antisymmetric matrix with $\epsilon$
blocks on the diagonal while the rest of the matrix elements being zero.
A generator $G$ of an $SP(2n)$ algebra in the spinor representation
obeys the condition
\beq
s G^T s = G \;\;\; s^2= -1.
\eeq
where the index $T$ means a transposition.

We now proceed to examine {\em massive} fermions in an external field.
In this model we have a left handed chiral fermion
$SU(2)$ doublet $q_L$ and a pair of right handed singlet fermions
which can be combined into a doublet $q_R$.
This is necessary for the introduction of fermionic masses.
The problem at hand now is how to generalize the definition of the
chiral fermionic determinant for the massive case.
We find it convenient and natural to define it as a square root of the
fermionic functional integral for two fermionic $(q_L^i,q_R^i)$,
$i=1,2$, multiplets with the same \mm \ $M$.
We follow, to this end, the recipe of ref. \cite{mah}:
we introduce a copy of the fermion system coupled to the same
external gauge field
and make a vectorization of the model to get a generalized
Dirac \op .

Let us consider a left-handed fermion doublet coupled to the $SU(2)$
gauge field and its copy
\beq
q_L = \left( \begin{array}{c} u\\ d\end{array}\right)_L, \qquad
q'_L = \left( \begin{array}{c} u'\\ d' \end{array} \right)_L,
\eeq
with associated weak singlet right-handed states as $u_R,\, d_R,\, u'_R,
\,d'_R$.

When the weak doublets are considered in pairs it is possible to pass to the
pure vector interaction of fermions with $W$ bosons. To achieve
this we introduce
instead of fermion fields $q'$ the charge-conjugated fields:
\beq
\tilde{q}'_R =\,\s C\bar{q}'_L = \left( \begin{array}{c} C\bar{d}'_L\\
-C\bar{u}'_L \end{array}\right), \qquad \tilde{q}'_L=\s
Cq'_R=\left(\begin{array}{c} C\bar{d }'_R \\ -C\bar{u}'_R \end{array}\right).
\eeq
Here $\s=i\sigma_2$ acts on the isotopic indices, $C$ is the
charge-conjugation matrix. By introducing
\beq
\begin{array}{lll} \psi=\psi_L+\psi_R,        &  \psi_L=q_L,   &
\psi_R=C\s\bar{q}'_L \\  \eta=\eta_R+\eta_L, & \eta_R=q_R, &
\eta_L=C\s\bar{q}'_R, \end{array}
\eeq
it becomes obvious that both components
of $\psi,\, \psi_L$ and $\psi_R$, are weak doublets while $\eta_R$ and
$\eta_L$ are singlets.
Therefore only the $\psi$ field has a vector gauge interaction:
\beq
L_W=i\p \nabla \hspace{-.65em}/
\psi + i\e\nn\eta, \qquad  \nabla \hspace{-.65em}/
=\gamma_\mu \nabla_\mu=\gamma_\mu (\partial_\mu -iW_\mu).
\eeq
Clearly the mixing of the quark and antiquark fields in eq.(2.15) is very
unnatural with respect to colour and electric charge (weak hypercharge).
As those interactions are
irrelevant to our problem at hand from now on we only keep the Yukawa
couplings:
\beq
-L_Y=h_u
\bar{q}_{Li}\s_{ij}u_R\r^*_j+h_d\bar{q}_{Li}d_R\r^i+H.c.+(u,d,h_u,h_d\to
u',d',h'_u,h'_d).
\eeq
We give here the Yukawa terms for different mass
matrices for two fermionic doublets.
This is convenient for explanation of properties of $D=5$ Dirac operator
which will be introduced below.
Actually for a discussion of the global anomaly we should put
$M= M' .$
Here the Higgs field is $\r^i=(\r^+,\r^0)$ and the fermion masses are
given by $m_u=h_uv/\sqrt{2}, \; m_d=h_dv/\sqrt{2},\; m'_u=h'_uv/\sqrt{2},\;
m'_d=h'_dv/\sqrt{2}$.

By using the fields $\psi$ and $\eta$ of eq.(2.5)
one can rewrite the Lagrangian
(2.7) as follows:
\beq
-L_Y =\p_L M\eta_R +\e_R M^+\psi_L-\p_R\s M'^*\s\eta_L-\e_L\s M'^T
\s\psi_R,
\eeq
where the mass matrix $M(x)$ is given by
\beq
M(x)\; =\; \left( \begin{array}{cc}  h_u\r^0(x)^*,  &   h_d\r^+(x)\\
-h_u\r^+(x)^*,  &  h_d\r^0(x)  \end{array} \right).
\eeq
$M'$ is taken from $M$ by substituting $h_u,h_d\to h'_uh'_d$.

We now make a Euclidean rotation upon which
the fermion fields rotate into
\beq
\psi,\, \eta\; \to\; \psi,\, \eta,   \qquad  \p,\, \e\; \to\; -\,i\psi^+,\,
-\,i\eta^+.
\eeq

The mass terms in $L_Y$ contain now the bilinear
combinations of the fields with the same chirality.
The Lagrangian reads now as follows
\beq
L = L_W + L_Y,
\eeq
\beq
L_W  =  -i\psi^+ \nabla \hspace{-.65em}/ \psi\, -\, i\eta^+\nn\,\eta ,
\eeq
$$L_Y  =  -i\psi^+_RM\eta_R - i\eta^+_L M^+\psi_L +i\psi^+_L\s M'^*\s\eta_L
+i\eta^+_R\s M'^T\s\psi_R.$$

We combine them into the Dirac spinors $\psi$ and $\eta$
whereas $\psi$ is an $SU(2)$ doublet and $\eta$ is a couple of $SU(2)$
singlet spinors.
As a result we get the \op \ in the fermionic kinetic term
as before for the particular case where $M=M'$.
This \op \ acts in the space of pairs $(\psi ,\eta)$ and reads as
\beq
T(M) = \left( \begin{array}{cc}  \nabla\hspace{-.75em}/ &
MR - \epsilon M^* \epsilon L\\
M^+ L - \epsilon M^T \epsilon R & \partial\hspace{-.55em}/
\end{array} \right),
\eeq
where $L(R) = (1+(-)\gamma_5 )/2$.

In refs. \cite{mah}
we restricted ourselves to the case of equal fermion masses
$(h_u=h_d)$.
In this case $M=-\epsilon M^* \epsilon$ with $M$ being
proportional to an element of the $SU(2)$ group.
This assumption was necessary with regard to a definition of an
appropriate chiral fermionic determinant and a $D=5$ antihermitian
euclidean \dop .
In this present paper we consider the case of an arbitrary mass matrix $M .$
It is easy to see that the \dop \ $T(M)$
has in general complex eigenvalues.
Hence when we continuously change the external gauge and Higgs fields
the levels move along the complex plane.
The non-zero eigenvalues of the $D=4$ \dop \ are paired
because of anti-commuting of $\Gamma_5$ with $T(M)$,
where
$$\Gamma_5 = \left( \begin{array}{cc} \gamma_5 & 0 \\
0 & -\gamma_5 \end{array} \right) .$$
Indeed for each non-zero eigenvalue $\lambda$ with the eigenfunction
$\psi_{\lambda}$ the function $\Gamma_5 \psi_{\lambda}$ corresponds
to the eigenvalue $- \lambda .$
As a result of such a pairing there exists an index theorem
for the \op \ $T(M)$
\beq
n_{{\rm Re} \alpha \lambda = 0} = {\rm topological \;\; invariant},
\eeq
where $\alpha$ is an arbitrary non-zero complex number and
$n_{{\rm Re} \alpha \lambda = 0}$ is a number of corresponding
eigenvalues,
and also
\beq
n_{\lambda = 0} = {\rm topological \;\; invariant}.
\eeq

It is now straightforward to define the square root of the determinant
of $T(M)$ as a product of the eigenvalues with
positive real part (${\rm Re} \lambda > 0 $)
for generic gauge and Higgs field configurations.
We thus reach to the following definition for the chiral fermionic
determinant
\beq
\Delta_{ch} = (det \; T(M))^{1/2}.
\eeq
The existence of Witten's anomaly implies that
an odd number of eigenvalues change their sign under a continuous
topologically non-trivial deformation of the external fields.
The conventional level crossing phenomenon would necessitate such
an exchange of fermionic levels to occur through the simultaneous
vanishing of the real and imaginary eigenvalues at some value of the
external fields. Alas in the space of all possible paths the
eigenvalues may choose to follow this one is truly of measure
zero. Indeed there is an infinitude of pathways where fermionic levels may
continuously circle around the origin ($\lambda=0$) of the complex plane
never in fact crossing each other (see fig.1). In this way the
adiabatic approximation which is assumed to be
valid is fully justified. We believe this to be the machanism by which
the pairs of fermionic levels exchange their sign. Level circling must
accordingly be equivalent to the presence of an odd number of zero
modes of the $D=5$ Dirac operator whose existence we proceed to
demonstrate.

In order to generalize our discussion of \lc \ we now define an
appropriate $D=5$ \op \ as
\beq
\hat{D} =
\gamma_5 \left( \begin{array}{cc}  \nabla_t  &  0\\ 0 &
-\partial_t \end{array} \right) +P\hat{T} =
\left( \begin{array}{cc} \gamma_5 \nabla_t + \nabla\hspace{-.75em}/ &
M R - \epsilon M^* \epsilon L\\
-M^+ L + \epsilon M^T \epsilon R &
-\gamma_5 \partial_t - \partial\hspace{-.55em}/ \end{array}
\right).
\eeq
The matrix $P$ is given by
$$P =
\left( \begin{array}{cc} 1 & 0\\ 0 & - 1
\end{array} \right) .$$

For the case of equal fermionic masses
the \op \ $\dd$ satisfies the following reality condition \cite{mah}
\beq
C\epsilon \gamma_5 \hat{D}^* C \epsilon \gamma_5 = \hat{D}.
\eeq
This is an important property that was used in refs. \cite{mah}
in the analysis of level crossing for the case of equal up and down
quark masses.
In effect there is a pairing up of all of the non-zero eigenvalues
in the spectrum of the \dop .
This furthermore implies that the number of its zero modes, if they exist,
is a topological invariant mod 2.

For a case of arbitrary Yukawa couplings this property is no more valid.
But we are still able to prove a pairing of non-zero levels
due to the antisymmetry property of the Dirac \op .
We have indeed
\beq
C\epsilon \gamma_5 P \hat{D}^T C \epsilon \gamma_5 P = - \hat{D}.
\eeq
This operation actually corresponds to a permutation of two identical
fermionic systems, $q \leftrightarrow q'$ and $M \leftrightarrow M' .$
Thus the antisymmetry of the \dop \ is not accidental but rather the
the consequence of the doubling used to make a well defined
fermionic path integral which is {\it a priori}
not the case for chiral fermions.

Notice that the antisymmetry of the \dop \ is well defined.
This is due to the reality of the representations of the group
$O(5) \times SP(2n)$. As a result the
transposed \dop \ along with an appropriate
conjugation with a constant matrix falls again in the same representation.
The pairing of levels can thus be formally proved as follows:

Let us consider the charachteristic
polynomial $\det (\lambda -\hat{D})$.
Because of antisymmetry of the \op \ $\hat{D}$ we have an identity
\beq
\det (\lambda -\hat{D}) = \det (\lambda +\hat{D}).
\eeq
Hence the non-zero eigenvalues of $\hat{D}$ are paired $(\lambda,
-\lambda) .$
As a consequence we see that the number of zero levels of
$\hat{D}$ is invariant modulo 2.
Moreover we have a generalization of the index theorems as given above
for the $D=4$ \dop \ (eqs. (2.25), (2.26)).

That means that there is an odd number of zero modes for
the $D=5$ \dop \ since it is the case when the masses of fermions are
equal.

Below we generalize an explicit construction of the \zm s in terms
of \zm s for massless \dop .

The solution $\Psi_0$ to the \zm \ equation
\beq
\hat{D}\Psi_0 = 0
\eeq
can be chosen real such that
\beq
C\epsilon \gamma_5 \Psi_0^* = \Psi_0.
\eeq
Notice that such a solution is not of any
definite chirality.
We may now represent $\Psi_0$ as a superposition of
an $SU(2)$ doublet
$\psi$ and two $SU(2)$ singlets $\chi_1$ and $\chi_2$, i.e.
$\Psi_0 = (\psi, \chi)$ with $\chi = (\chi_1 ,\chi_2)$.
The zero mode equations then read as
\beq
\dd \psi + (M R -\epsilon M^* \epsilon L )\chi = 0, \qquad
(M^+L - \epsilon M^T \epsilon R) \psi + \dd_0 \chi = 0,
\eeq
where
\beq
(\gamma_5 \nabla_t + \nabla\hspace{-.65em}/ ) = \dd, \qquad
(\gamma_5 \partial_t + \nn ) = \dd_0.
\eeq
By eliminating $\chi$ from eqs.(2.34)
we get
\beq
\dd \psi = (M R - \epsilon M^* \epsilon L) (1/ \dd_0)( M^+ L -
\epsilon M^T \epsilon R) \psi.
\eeq
Let us assume that there is exactly one \zm \ of the \op \ $\dd .$
We want to find the massive \zm \ wave function,
and moreover express the solution of the above equation
in terms of the \zm \ wave function $\psi_0$
for the massless case.

We search for a solution iteratively by starting with $\psi_0$
as a zero
level approximation.
The first level correction satisfies the following equation
\beq
\dd \psi^{(1)} = (M R - \epsilon M^* \epsilon L) (1/ \dd_0)( M^+ L -
\epsilon M^T \epsilon R) \psi_0 .
\eeq
The integrability \cn \ is deduced from the orthogonality
of the left hand side of eq.(2.36) to $\psi_0$
\beq
a_0 = \int d^5 x  \psi^+_0 (M R - \epsilon M^* \epsilon L) (1/ \dd_0)( M^+ L -
\epsilon M^T \epsilon R)\psi_0 = 0.
\eeq
Indeed if we make a transposition in eq.(2.37) and susequently take into
account the antisymmetry \cn  \ for the \op \
$\dd$
\beq
C\epsilon \gamma_5 \dd^T_0 C \epsilon \gamma_5  = - \dd_0 ,
\eeq
and the reality \cn \ (2.32)
we get that $a_0 = -a_0 =0$.

We may now solve eq.(2.36)
\beq
\psi^{(1)} = \frac{1}{\dd} (M R - \epsilon M^* \epsilon L) (1/ \dd_0)( M^+ L -
\epsilon M^T \epsilon R) \psi_0.
\eeq
This is a well defined expression on the basis of eq.(2.37).
The exact solution to eq.(2.35) is given by
\beq
\psi = \psi_0 + \dd \frac{1}{\dd^2 + \alpha P -
N \dd_0^{-1} \tilde{N} \dd } \\
N \frac{1}{\dd_0} \tilde{N} \psi_0,
\eeq
where $P$ is the \zm \ subspace projector and $\alpha$ a regularizing
parameter. We have denoted
\begin{equation}
N = M R - \epsilon M^* \epsilon L , \;\;\;
\tilde{N} = M^+ L -\epsilon M^T \epsilon R .
\end{equation}
This expression is well defined and does not depend on
the regularizing parameter $\alpha$.
By substitution of (2.41) for $\psi$
into eq.(2.35)
we get the following equation
\beq
P \frac{1}{\dd^2 + \alpha P - N \dd_0^{-1}
\tilde{N} \dd} \; N\frac{1}{\dd_0} \tilde{N} \psi_0 =0.
\eeq
By using the block
matrix representation \cite{mah}
we find an equivalent form of eq.(2.43)
\begin{eqnarray}
\int d^5 x \psi_0^+ \left( N \frac{1}{\dd_0} \tilde{N} +
N \frac{1}{\dd_0} \tilde{N} (1-P)
\frac{1}{\dd  -
(1-P) N \dd_0^{-1} \tilde{N} (1-P)}\times \right. \nonumber \\
\left. \times (1-P) N \frac{1}{\dd_0} \tilde{N}\right)
\psi_0 =0.
\end{eqnarray}
The above expression is well defined.
It is certainly satisfied as a consequence of the antisymmetry
of the \op s $\hat{D},\dd_{0}$ and $\dd .$
This can be checked by a transposition similar to the case of eq.(2.38).

Let us now turn to the issue of \nty \ of $\psi$ in eq.(2.41).
Let us demonstrate that the wave function (2.41) is \nl . For that it
is sufficient to check that expression (2.41) decreases rapidly enough
at large distances $x\to \infty$.
This we can more conveniently do in the regular gauge for the \inst \
configuration.
At infinity we have that
\beq
D_\mu \to U \partial_\mu U^{-1},
M \to U M_0,
\eeq
where $U$ is an element of the $SU(2)$ group that corresponds to the
\inst \ configuration and $M_0$ is a constant matrix.
By a direct substitution of expressions (2.45)
back into eq.(2.41)
we easily find that
\beq
\psi_L \propto U\frac{\partial^2}{\partial^2 - M_0^2} U^{-1} \psi_{0L}.
\eeq
It becomes obvious from the above that
$\psi_{L}$ behaves better at infinity than $\psi_{0L}$
itself
a \nl \ wave function.
Indeed we trivially deduce from eq.(2.46) that
\beq
\psi_{L}\leq U\frac{1}{x^2 M_0^2} U^{-1} \psi_{0L}.
\eeq
Hence $\psi_{L}$ is \nl .
It is easy to see that this wave function decreases more rapidly at
infinity provided that the \mm \ $M$ is an asymptotically covariant
constant similar to the case of the $D=4$
\inst \ $+$ Higgs \con s.
This necessitates that the gauge \f \ is asymptotically a pure gauge
$A_{\mu} \propto U
\partial_{\mu} U^{-1}$ with the \mm \ being $M \propto UM_0$
at infinity where $U$ is an element of the $SU(2)$ group and $M_0$ is a
constant.
In this case the wave function (2.41) is normalizable.
It is to be emphasized at this point that by changing $M$ from zero to
a nonzero value we can obtain additional zero modes but always in pairs.
It is of some significance to note that this \cn \ implies that
in the presence of Yukawa interactions the interpolation between
two $D=4$ \con s of gauge fields $A_{\mu}$ and $A^U_{\mu}$
which is given by a $D=5$ gauge \f \ \con \ should also include
Higgs \f s.

We now turn to the case where the \op \ $\dd$ has a number of
normalizable zero modes $\psi_{0i}$, $i=1,...,N$.
These zero modes can be chosen to obey the reality condition
\beq
C \gamma_5 \epsilon \psi_{0i}^* = \psi_{0i}.
\eeq
The possibility of such a choice seems to be crucially necessary
for the following since allows to prove the integrabilty of the
equation for zero mode.

In order to solve eq.(2.36) we start the
iteration procedure with a linear combination
\beq
\psi_0 = c_i \psi_{0i}
\eeq
as a zero level approximation.
At the first step we get eq.(2.37).
The integrability condition reads now as
follows
for all $i$
\beq
\int d^5 x \psi^+_{0i} N \dd_0^{-1} \tilde{N} \psi_0 = 0.
\eeq
This means that the vector $c_i$ should be annihilated by the matrix
\beq
a^{(0)}_{ij} = \int d^5 x \psi^+_{0i} N \dd_0^{-1} \tilde{N}
\psi_{0j}.
\eeq
A solution of eq.(2.37) and hence of eq.(2.36)
exists only if the matrix $a^{(0)}_{ij}$
has zero eigenvalues.
By considering the Hermitian conjugate and the complex one of $a_{ij}$
we can easily check
that this matrix is antisymmetric but complex in general
in contrast to the case of equal fermionic masses \cite{mah}
\beq
a^{(0)}_{ij} = -a^{(0)}_{ji} .
\eeq
In general this matrix is non-zero and belongs to the
complexification of $O(N)$ algebra.
Let us  first consider the case of an odd number of $\psi_0$'s, i.e.
$N=2n+1$.
It is easy to see that the matrix $a^{(0)}_{ij}$ has at least one zero
eigenvalue.

In order to show that it is sufficient to consider the charachteristic
polynomial $\det (\lambda -a^{(0)})$.
Because of the antisymmetry of the matrix $a^{(0)}$ we have
\beq
\det (\lambda -a^{(0)}) = \det (\lambda +a^{(0)}).
\eeq
Hence the non-zero eigenvalues of $a^{(0)}$ are paired $(\lambda,
-\lambda) .$
Moreover we see that the determinant of $a^{(0)}$ is zero for any odd
dimension of $a^{(0)} .$
As a consequence we see that the number of zero levels of
$a^{(0)}$ is invariant modulo 2.
Therefore since for the real $a^{(0)}$ of odd number there is
an odd number of zero eigenvalues the same is correct for the
complex $a^{(0)} .$

In the case of even $N$, i.e. $N=2n$,
all eigenvalues of $a^{(0)}_{ij}$ are generically non-zero and
are arranged in
pairs $(i\lambda, -i\lambda)$ as above.

We get the zero level approximate
solution to eq.(2.36) by picking the eigenvector of $a^{(0)}_{ij}$ that
corresponds to the zero eigenvalue for the vector $c_i$ in eq.(2.49).
It is obvious from the above construction that the
'oddness' of the number of the fermionic
\zm s does not change.
In particular there is at least one normalizable zero mode
of massive fermion in the $D=5$ topologically nontrivial
configuration since
it has exactly one zero mode for massless ones.

The exact solution to eq.(2.36) is given by (2.41) where $P$ is the
projector onto the total \zm \ subspace.
We must be more careful here than with the case of the one massless \zm \
$(N=1)$ since in general the function $N \dd_0^{-1} \tilde{N}
\psi_0$ is not
orthogonal to the \zm \ subspace.
By the use of the block matrix representation it is easy to see that the
expression (2.41) obeys eq.(2.36) provided that the following holds true
for all $i$
\begin{eqnarray}
\int d^5 x \psi_{0i}^+ \left( N \frac{1}{\dd_0} \tilde{N} +
N \frac{1}{\dd_0} \tilde{N} (1-P)\times \;\;\; \right. \\
\left. \times \frac{1}{\dd  -
(1-P) N \dd_0^{-1} \tilde{N} (1-P)}
(1-P) N \frac{1}{\dd_0} \tilde{N}\right)
\psi_0 =0. \;\; \; \nonumber
\end{eqnarray}
The above expression is well defined and implies that the vector
$c_j$ should be the eigenvector that corresponds to the zero
eigenvalue of the following matrix
\begin{eqnarray}
a_{ij} = \int d^5 x \psi_{0i}^+ \left( N \frac{1}{\dd_0} \tilde{N} +
N \frac{1}{\dd_0} \tilde{N} (1-P)\times \;\;\; \right. \\
\left. \times \frac{1}{\dd  -
(1-P) N \dd_0^{-1} \tilde{N} (1-P)}
(1-P) N \frac{1}{\dd_0} \tilde{N}\right)
\psi_{0j}, \nonumber \;\;\;
\end{eqnarray}
i.e.
\beq
\sum_j \; a_{ij}\; c_j = 0.
\eeq
In the leading order in the Yukawa
couplings this \cn \ reproduces eq. (2.49).
It can be easily seen that it is equivalent to the \igty \
\cn \ analogous to eq.(2.33).
One may observe that the matrix $a_{ij}$ is
antisymmetric by using the reality \cn \ (2.32) and the antisymmetry
of the \op s $\dd$ and $\hat{D} .$
By repeating here the procedure we followed for the leading
approximation we find that the matrix $a_{ij}$
has exactly an odd number of zero eigenvalues for odd $N=2n+1$
and correspondingly an even number of zero eigenvalues for even
$N=2n$.
In the latter case the zero eigenvalues can be absent.
Hence there is an invariance for the odd/even number of fermionic
zero modes under smooth deformations of the \mm \ $M .$

At this point our arguments that generalize Witten's observation of
an $SU(2)$ \an \ for odd number of massless fermion doublets in the
presence of Yukawa
interactions are complete.
The existence of the \zm \ for massive $D=5$ \dop \ was demonstrated
for the general case of a nondegenerate \mm .

\subsection*{Conclusions}
In the present paper we looked at the Witten $SU(2)$ global anomaly for
the case of an odd number of $SU(2)$ Weyl doublets
in an $SU(2)$ gauge theory with arbitrary Yukawa couplings and fermion
masses. The $D=4$ Dirac operator appears to be antisymmetric and
nonhermitean. As a consequence its spectrum of nonzero eigenvalues
appears in pairs with opposite sign and is complex. We showed explicitly
the existence of an odd number of normalizable zero modes for the $D=5$
Dirac operator. Their presence is equivalent to a level exchange
phenomenon, level "{\em circling}", between the fermionic level pairs under
continuous deformations of the external gauge field. This is a direct
generalization of the well established level crossing phenomenon and
applies to all $SP(n)$ gauge theories with  arbitrary Yukawa couplings
and fermion masses coming in an odd number in its spin representation.

\subsection*{Acknowledgments}
One of us A.J. acknowledges the high energy group at NBI for its
hospitality. The present work was supported in part by a NATO grant
GRG 930395. M.A. is grateful to the Carlsberg Foundation for financial
support.

\newpage
\subsection*{Figure Caption}
\vskip 6pt
Fig 1. The level circling phenomenon is depicted for an odd number of
fermionic level pairs $(\lambda,-\lambda)$ on the complex plane. Under
continuous gauge field deformations the lowest lying pair
switches sign by circling
around the origin.

\end{document}